\documentclass[11pt,twocolumn]{article}
\usepackage[dvipdfmx]{graphicx}
\expandafter\let\csname equation*\endcsname\relax
\expandafter\let\csname endequation*\endcsname\relax
\usepackage{amsmath}
\usepackage{physics}
\usepackage{textcomp}
\usepackage{color}
\usepackage{ulem}
\usepackage[margin=17mm]{geometry}
\begin{document}
\setlength{\columnsep}{8mm}
\twocolumn[%
\noindent
\textbf{\Large
{Monochromatic Electron Emission  from Graphene-Insulator-\\ 
Semiconductor-Structured Electron Source Utilizing Interference Effects}
}

\vspace{3mm}
\noindent
{\large
Takao Koichi$^{1,2}$, 
Shogo Kawashima$^1$, 
Hiroshi Miyake$^1$, 
Satoshi Abo$^1$, 
Fujio Wakaya$^{1*}$,\\
Masayoshi Nagao$^2$, 
Katsuhisa Murakami$^2$
}

\vspace{2mm}

\noindent
$^1$School/Graduate School of Engineering Science, The University of Osaka, 
Toyonaka 560-8531, Japan.

\noindent
$^2$National Institute of Advanced Industrial Science and Technology, 
Tsukuba 305-8568, Japan.

\vspace{2mm}

\noindent
$^*$E-mail: {wakaya.f.es@osaka-u.ac.jp}

\renewcommand{\abstractname}{}

\vspace{5mm}

\noindent
The graphene-insulator-semiconductor-structured electron source has garnered significant attention due to its high electron emission efficiency and highly monochromatic electron emission. 
Graphene, with its c-axis orientation and well-defined interlayer spacing, exhibits electron interference effects that can influence the properties of emitted electrons. 
In this work, motion of an electron wave packet is numerically calculated to discuss the 
energy spread of the zero-order and first-order diffracted electron waves by mono- and multilayer graphene. 
It is found that the effects of multiple reflections of electron between the layers broaden the energy spread 
especially for the incident energy of 13.4 eV, and that highly monochromatic electron emission can be achieved by using diffracted electron wave with a small aperture.   \\

\vspace{-3mm}

\noindent
\textbf{keywords:} Graphene, GIS electron source, Simulation

\vspace{10mm}

]

\section{Introduction}

A planer-type electron source offers several advantages, such as low operating voltage, operation in low vacuum \cite{MIM1,MIM2}, and a small divergence angle of emitted electrons \cite{shimawaki}. A planer-type electron source typically consists of three stacked layers, \textit{i.e.} metal/insulator/metal (MIM) \cite{MIM1,MIM2,MIM0} or metal/insulator/semiconductor (MIS) \cite{yokoo1993,yokoo1994,yokoo1996,yokoo2006}. The thicknesses of the top and insulator layers are typically 3 nm and 10 nm, respectively. When a negative bias voltage ($\sim -10$ V) is applied to the top layer relative to the substrate, the Fowler-Nordheim (FN) tunneling \cite{fowler,nordhiem} occurs. The tunneling electrons are then accelerated toward the top layer by the high electric field in the insulator. Electrons with an energy of $\sim 10$ eV, incident vertically on the top layer, pass through it and are emitted into the vacuum if their energy exceeds the work function of the top layer material.  

A high emission efficiency of 48.5\% \cite{100,321,furuya,485} has been achieved using a graphene-insulator-semiconductor (GIS)-structured electron source, which is more than four orders of magnitude greater than that of conventional MIM and MIS devices. Additionally, highly monochromatic electron emission has been reported by using hexagonal boron nitride ({\it h}-BN) as the insulator layer in the GIS electron source \cite{igari}. The energy spread of the emitted electron beam from the graphene/{\it h}-BN/n-Si electron source is 0.18 eV, which is smaller than that from cold-field-emission cathodes \cite{Ta03}. The authors attribute these excellent properties to the suppression of inelastic scattering in both the topmost and insulator layers.

\begin{figure*}[tb]
  \vspace{0pt}
  \centering
  \includegraphics[scale=0.6]{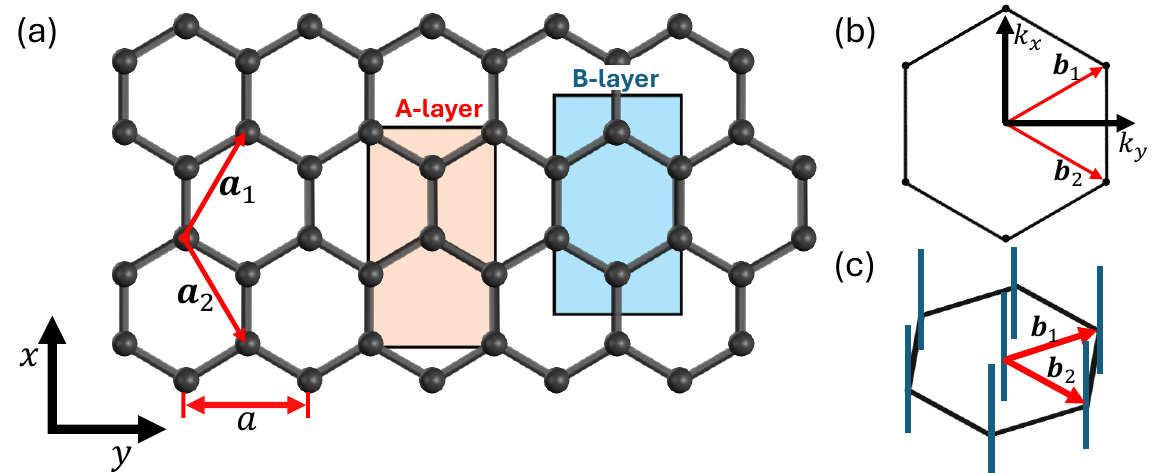}
  \caption{(a)Monolayer graphene structure. $\vb*{a}_1$ and $\vb*{a}_2$ are primitive lattice vectors. 
  $a = \abs{\vb*{a}_1} = \abs{\vb*{a}_2} = 2.46$ \AA\ is the lattice constant.
 The A- and B-layers 
 in the conventional unit cell  
 for multi-layer graphene 
 that is adopted in the present simulation are 
 also 
 shown.
  (b)The reciprocal lattice of monolayer graphene. 
  $\vb*{b}_1$ and $\vb*{b}_2$ are primitive reciprocal lattice vectors.
  (c)3D reciprocal structure of monolayer graphene with Bragg scattering rods.}
  \vspace{0pt}
  \label{fig:structure123}
\end{figure*}

The graphene deposited on the insulator is oriented along the $c$-axis \cite{castro2009electronic}, exhibiting a distinct interlayer spacing. These characteristics may significantly influence the properties of the emitted electron beams. The energy-filtering effect due to multiple reflections of graphene layers, along with the diffraction characteristics of electron waves at the graphene, could potentially enhance the monochromaticity of the emitted electrons. This discussion, however, is absent from the current literature. The aim of this work is to propose a novel method to achieve highly monochromatic electron emission by utilizing the effects of multiple reflections or diffraction at graphene. To investigate these effects, the motion of an electron wave packet with an energy of several tens of eV, incident vertically on the graphene, is simulated. 
%
In this work, the electron transmission properties of single and several-layer graphene are calculated to discuss the properties of electron emitted from GIS-structured electron source. 
Such a simplification should be allowed because 
the scattering and diffraction processes at the topmost graphene layer contribute most significantly to the property of electron emitted from GIS-structured electron source.  
Since this work focuses on highly monochromatic electron emission from GIS-structured electron source, detailed aspects of graphene-electron interactions, 
such as the incident angle dependence, twisted structures, and scattering intensity, 
are not discussed\cite{PENG1999625,Wicki,LATYCHEVSKAIA201946}.
%
%
The results show that the monochromaticity of the emitted electrons improves by exploiting electron diffraction at graphene.

\section{\label{sec:level1} Simulation Method}

Figure~\ref{fig:structure123}(a) shows the crystal structure of monolayer graphene. In the hexagonal lattice of monolayer graphene, the primitive lattice vectors $\vb*{a}_1$ and $\vb*{a}_2$ are given by 
\begin{equation}
  \vb*{a}_1 = (\frac{\sqrt{3}a}{2},\frac{a}{2},0),\,\,
  \vb*{a}_2 = (\frac{\sqrt{3}a}{2},-\frac{a}{2},0),
  \label{}
\end{equation}
where $a = \abs{\vb*{a}_1} = \abs{\vb*{a}_2} = 2.46$ \AA\ is the lattice constant \cite{castro2009electronic,kavitha2016graphene,yang2018structure}. The primitive reciprocal lattice vectors $\vb*{b}_1$ and $\vb*{b}_2$ are
\begin{equation}
  \vb*{b}_1 = (\frac{2\pi}{\sqrt{3}a},\frac{2\pi}{a},0),\,\,
  \vb*{b}_2 = (\frac{2\pi}{\sqrt{3}a},-\frac{2\pi}{a},0).
  \label{}
\end{equation}
The reciprocal lattice is also hexagonal \cite{mccann2012electronic,McCann_2013}. Figures~\ref{fig:structure123}(b) and (c) depict the reciprocal lattice. The reciprocal lattice structure consists of infinite rods extending along $\vb*{\hat{k}}_z$, as the monolayer graphene lattice is a two-dimensional plane perpendicular to $\vb*{\hat{z}}$ \cite{meyer2007roughness,meyer2007structure,sung2019stacking}. The Bragg scattering rods are spaced by 
$\abs{\vb*{b}_1} = \abs{\vb*{b}_2} = \frac{4\pi}{\sqrt{3}a} = 2.95$ \AA$^{-1}$.

The first-order reciprocal lattice vectors are defined as
\begin{equation}
  \vb*{g}_1 = \frac{4\pi}{\sqrt{3}a}(\pm 1,0,0)\ \rm{or}\ \frac{4\pi}{\sqrt{3}a}(\pm 1/2,\pm \sqrt{3}/2,0).
  \label{}
\end{equation} 
The magnitude of $\vb*{g}_1$ is the same as the primitive reciprocal lattice vectors $\vb*{b}_1$ and $\vb*{b}_2$, \textit{i.e.} 
$\abs{\vb*{g}_1} = \abs{\vb*{b}_1} = \abs{\vb*{b}_2} = 2.95\,\text{\AA}^{-1}$. The electron energy corresponding to $\vb*{g}_1$ is  
\begin{equation}
  E_{\vb*{g}_1} = \frac{\hbar^2\abs{\vb*{g}_1}^2}{2m} = 33.14\ \rm{eV} ,
\end{equation}
where $\hbar$ is the Dirac's constant and $m$ is the electron mass. First-order diffraction occurs when an electron with energy greater than 33.14\,eV passes through the graphene with momentum along the $z$ direction. The first-order diffracted electron has a momentum component perpendicular to the $z$ direction, with a velocity of $\vb*{v}_{\vb*{g}_1} = \hbar \vb*{g}_1 / m$.

In this work, the multilayer graphene is assumed to be ABA stacked, as the ABA-stacked configuration is generally more stable than other graphene structures \cite{ABA1,ABA2}. The distance between the graphene layers is 3.35 \AA. 
The A- and B-layers 
in the conventional unit cell  
used in the numerical calculation in the following section are shown in Fig.~\ref{fig:structure123}(a) in the same $x$-$y$ plane. In ABA-stacked multilayer graphene, the primitive lattice vector in the $z$ direction is $\vb*{a}_3 = (0,0,c)$, where $c = 6.7\ \text{\AA}$ \cite{Gray2009CrystalSO}. The corresponding primitive reciprocal lattice vector is 
$\vb*{b}_3 = \frac{2\pi}{c}(0,0,1)$, with $\abs{\vb*{b}_3} = \frac{2\pi}{6.7} = 0.938\ \text{\AA}^{-1}$.

\begin{figure}[t]
\centering
\includegraphics[width=8cm]{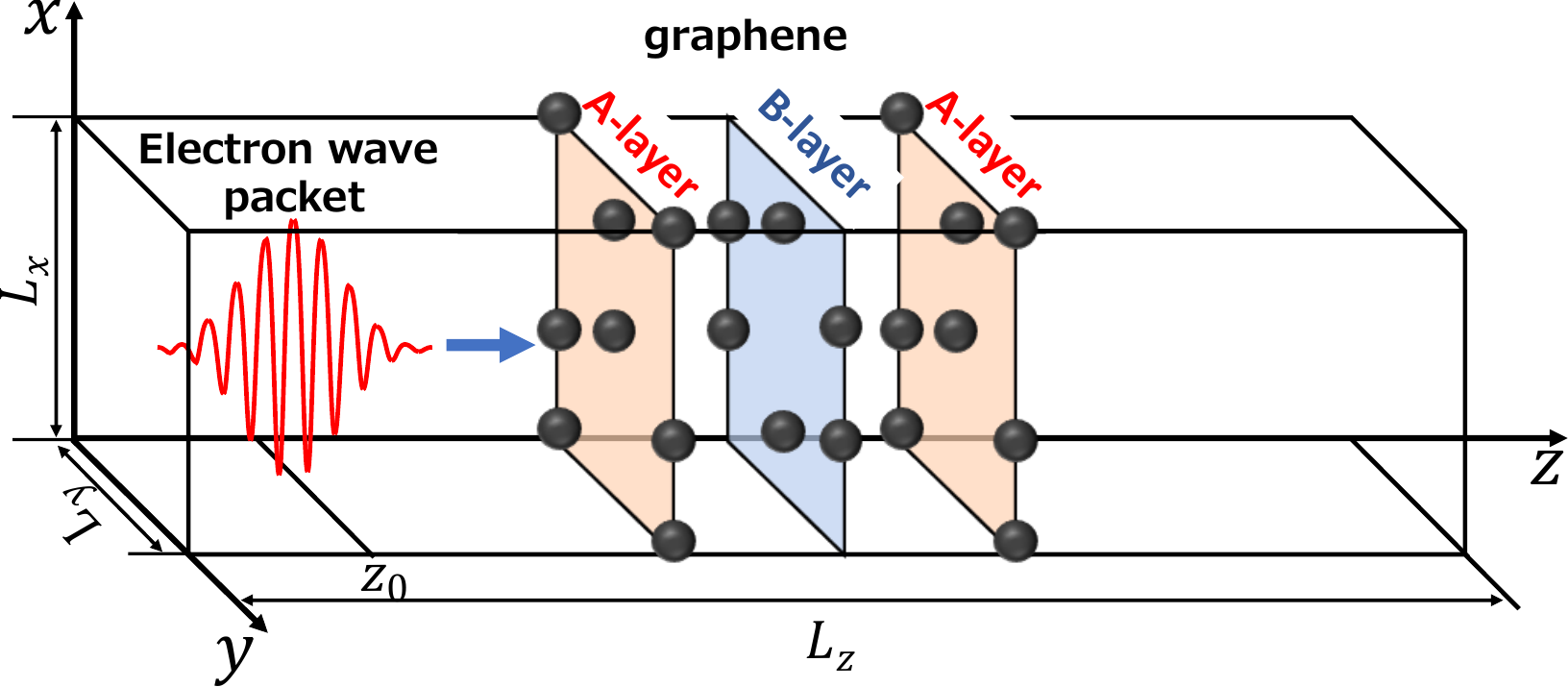}
\caption{
Illustration of three-dimensional space used in the simulation.  
The unit cell in the $x$-$y$ plane is defined as $L_x \times L_y$, 
and is assumed to be periodic.    
The graphene is illustrated as ABA-stacked three-layer one, 
although the simulation is performed not only for three-layer one 
but also single or multi-layer one other than three-layer.
}
\label{grapheneh}
\end{figure}

Figure~\ref{grapheneh} shows an illustration of the three-dimensional space used in the present simulation. The boundary conditions of the unit cell, with lengths $L_x$ and $L_y$, are assumed to be periodic. The length in the $z$ direction is $L_z$. Complex absorbing potentials (CAPs) \cite{CAP1,CAP2,CAP3} are placed at both boundaries in the $z$ direction. 

\begin{figure*}[tb]
  \vspace{0pt}
  \centering
  \includegraphics[width=17cm]{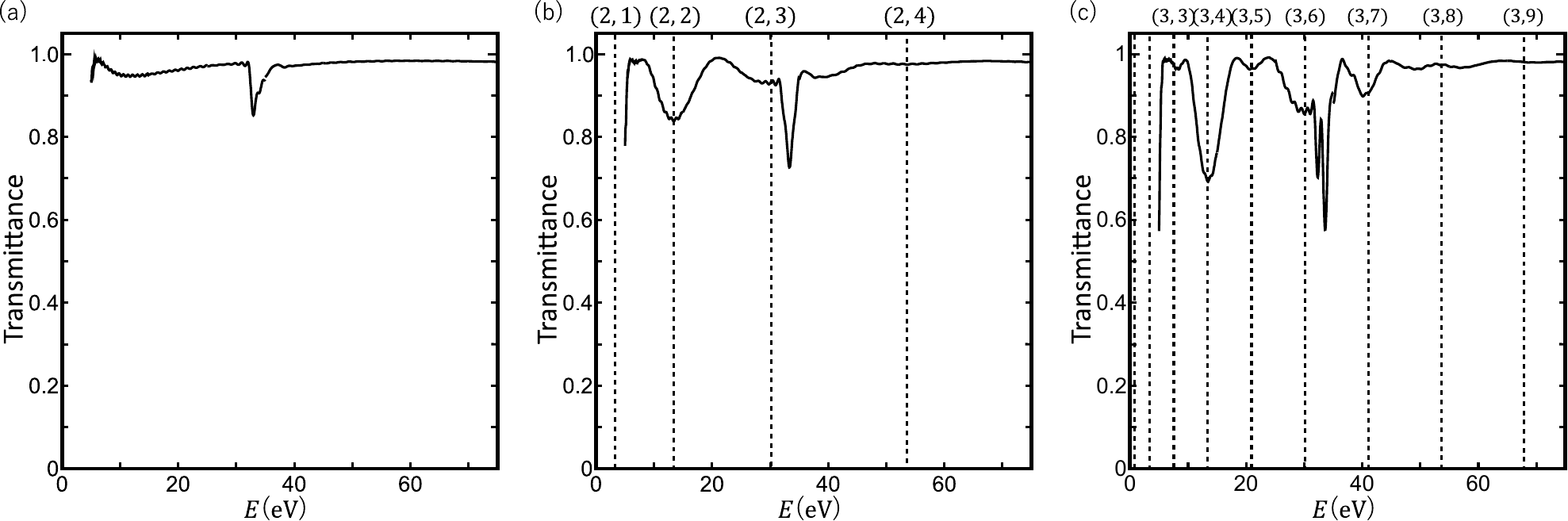}
  \caption{Transmittance as a function of electron energy using monolayer to trilayer graphene.
  $\alpha_0 = 10 $\ \AA, $E_0 = 10,20,\cdots70 \,\mathrm{eV}$.  
  At the dotted lines,  $\lambda /2 \times n = 3.35 \times (l-1)$ is satisfied, where $\lambda$ is wavelength, $l$ is the number of graphene layers, $n$ is positive integer.
  The numbers above the graphs represent ($l,n$).}
  \vspace{0pt}
  \label{RT1}
\end{figure*}

\begin{figure}[tb]
  \vspace{0pt}
  \centering
  \includegraphics[scale=0.6]{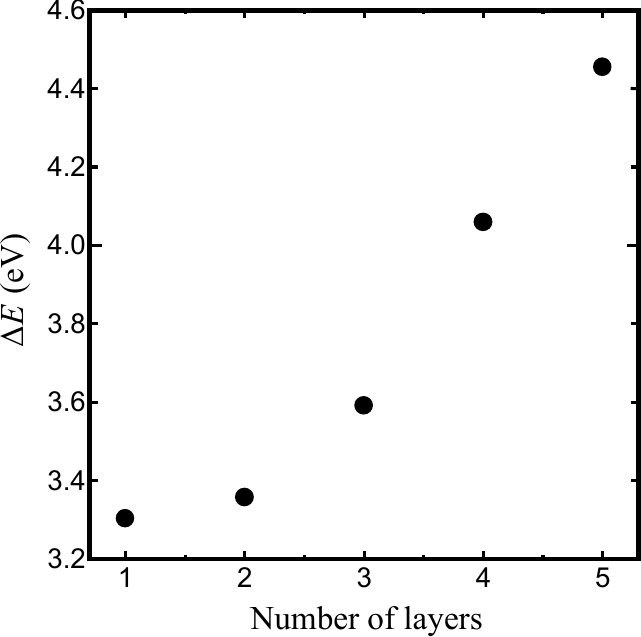}
  \caption{FWHMs of the EEDs transmitted through monolayer to five-layer graphene. 
  $E_0=$13.4\ eV, $\alpha_0 = 10$\ \AA.}
  \vspace{0pt}
  \label{13.4}
\end{figure}

The initial Gaussian wave packet is assumed to be
\begin{equation}
\psi_0 = \frac{1}{\sqrt{L_xL_y}}\left(\frac{2}{\pi\alpha_0^2}\right)^{\!\!\frac{1}{4}} e^{ik_0z-\frac{(z-z_0)^2}{\alpha_0^2}},
\label{eq:initial}
\end{equation}
where $\alpha_0$ is related to the wave-packet size, $z_0$ is the center of the Gaussian function, $k_0=\sqrt{2mE_0}/\hbar$ is the wavenumber of the incident electron, $m$ is the electron mass, and $E_0$ is the electron energy of the plane wave. The electron wave packet initially propagates in the $+z$ direction. The wave function is evolved from the initial one by solving the Schr\"odinger equation. The graphene potential $V(\vb*{r})$ is obtained using the Thomas-Fermi approximation and the screened Coulomb potential, with the electron density in graphene calculated by density functional theory \cite{QE}. In the numerical simulation, the time evolution of the wave packet is calculated by
\begin{equation}
  \psi (\vb*{r},t + \Delta t) = \exp(-\frac{iH\Delta t}{\hbar})\psi(\vb*{r},t).
  \label{eq:timev}
\end{equation}

The wave function $\psi(\vb*{r}, t)$ is transformed from the time domain into energy space as 
\begin{equation}
  \Phi (\vb*{r},E) = \frac{1}{2\pi}\int \psi(\vb*{r},t)e^{i\frac{Et}{\hbar}} dt.
  \label{f2,4}
\end{equation}
Using the energy-space wave function, the transmission coefficients is calculated as follows \cite{simulation_yan}. If $ E_{\vb*{g}_1} < E < E_{\vb*{g}_2}$, the transmitted plane wave can be written as
\begin{equation}
  \Phi (\vb*{r},E) = T_{E}e^{ik z} + \sum_{\vb*{g}_1} T_{E \vb*{g}_1}e^{i\vb*{g}_1 \cdot \vb*{\rho}}e^{ik_{\vb*{g}_1\perp} z},
  \label{eq:transmission}
\end{equation}
where $T_{E}$ 
and $T_{E \vb*{g}_1}$ 
are the transmission coefficients of the zero-order and first-order diffracted waves, respectively. 
The calculation plane perpendicular to the $z$ axis is placed at a position posterior to the graphene. 
The Fourier transform of $\Phi (\vb*{r},E)$, performed over the calculation plane, 
yields $T_{E}$ and $T_{E \vb*{g}_1}$ as 
\begin{equation}
  T_{E} = \int \Phi (\vb*{r},E) e^{-ikz} dx dy, 
  \label{eq:transmission0}
\end{equation}
\begin{equation}
  T_{E \vb*{g}_1}= \int \Phi (\vb*{r},E) e^{-i\vb*{g}_1 \cdot \vb*{\rho}}e^{-ik_{g_1\perp} z} dx dy.  
  \label{eq:transmission1}
\end{equation}

\section{Results and Discussion}

\subsection{Interlayer Interference Effect}

\begin{figure*}[tb]
  \vspace{0pt}
  \centering
  \includegraphics[width=11cm]{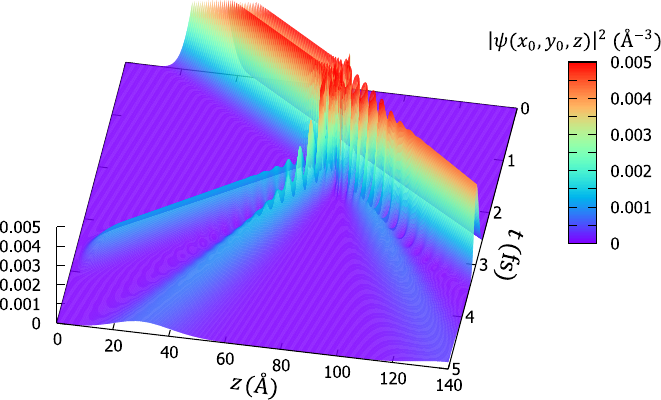}
  \caption{Electron wave packet with $E_0=40$ eV as a function of $z$ and time.
  The position $(x_0,y_0)$ is the center of the conventional unit cell.
  Monolayer graphene is placed at $z=85.7$ \AA.}
  \vspace{0pt}
  \label{E0=40}
\end{figure*}

\begin{figure}[tb]
  \vspace{0pt}
  \centering
  \includegraphics[scale=0.65]{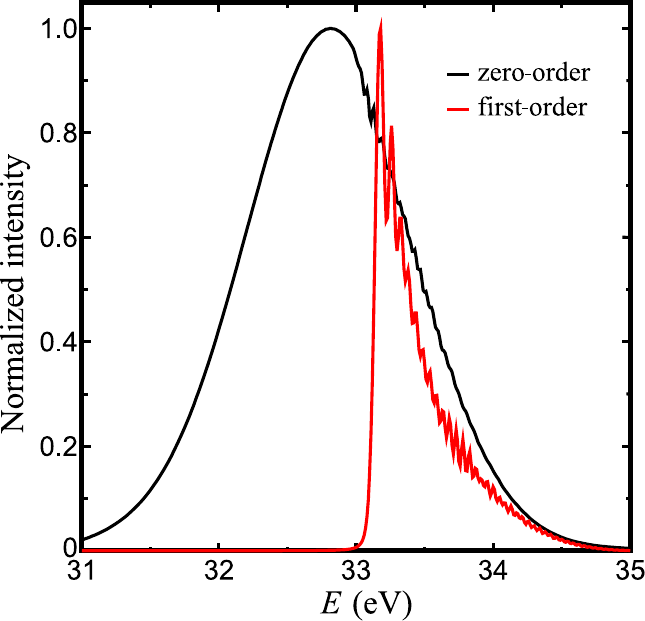}
  \caption{EEDs of the zero-order and  first-order diffracted waves
  after transmission through monolayer graphene with $E_0 = 33$ eV and $\alpha_0 = 40$ \AA.
  Black and red lines represent the zero-order and first-order diffracted waves, respectively.}
  \vspace{0pt}
  \label{E0=33}
\end{figure}

The transmittance $\abs{T_{E}}^2+\sum_{\vb*{g}_1} \abs{T_{E \vb*{g}_1}}^2$ is calculated at the calculation plane positioned posterior to the graphene, as described in Ref.~\cite{simulation_yan}. This plane is located approximately 50 \AA\ away from the graphene.
Figure~\ref{RT1} presents the transmittance as functions of electron energy for monolayer, bilayer, and trilayer graphene. The calculations were performed within an energy range of $5$ -- $75$ eV. 
These simulations neglect inelastic scattering\cite{Takao1,Miyauchi,LATYCHEVSKAIA2023113807}, leading to discrepancies between the calculated transmittance and experimental data \cite{485,Longchamp}. 
However, the calculated results in the present work can be used for 
analyzing how multiple reflections in graphene structures affect electron monochromaticity.
In the actual experiments, the intensity of the effects should be reduced to some extent. 

As shown in Fig.~\ref{RT1}(a), the transmittance for monolayer graphene remains nearly unity, except at $E \sim 33.14$ eV. This high transmittance is attributed to the absence of inelastic scattering effects in the present study. 
At $E \sim 33.14$ eV, the first-order diffracted wave propagates predominantly in the $x$-$y$ plane.
Consequently, this wave fails to reach the calculation plane positioned away from the graphene, resulting in a distinct reduction in transmittance at this specific energy.

For multilayer graphene, the transmittance exhibits several dips at specific electron energies, notably at $E = 13.4$ eV and $30$ eV. At $E = 13.4$ eV, the electron wavelength $\lambda$ is $3.35$ \AA, corresponding to the interlayer distance in graphene. Similarly, at $E = 30$ eV, $\lambda$ is $2.24$ \AA, which is two-thirds of the interlayer spacing. These wavelengths satisfy an interference condition described by the equation:
\begin{equation}
  \lambda /2 \times n = 3.35 \times (l-1),
  \label{eq:interference}
\end{equation}
where $\lambda = \frac{2\pi\hbar}{\sqrt{2mE}}$ is the electron wavelength, $l$ is the number of graphene layers, and $n$ is a positive integer. The dashed lines in Figs.~\ref{RT1}(b) and (c) indicate the electron energies satisfying Eq.~(\ref{eq:interference}). These findings strongly suggest that the dips in transmittance are attributed to the interlayer interference effect in multilayer graphene structures.

The transmittance shows a significant reduction at $E = 13.4$ eV, as seen in Fig.~\ref{RT1}, indicating that electrons at this energy are substantially affected by interlayer interference in graphene structures. Figure~\ref{13.4} illustrates the simulated full width at half maximum (FWHM) of the electron energy distributions (EEDs) transmitted through monolayer to five-layer graphene for $E_0=13.4$ eV and $\alpha = 10$ \AA. The FWHM of the EED increases as the number of graphene layers increases. The calculated energy spread suggests that the interlayer interference effect, regardless of the number of graphene layers, widens the energy distribution at $13.4$ eV. This implies that a gate voltage of 13.4 V should be avoided when aiming to realize a highly monochromatic electron source. Moreover, there are no other electron energies that can achieve highly monochromatic electron emission using multiple reflections in graphene structures. This is due to the absence of specific energies with significantly higher transmittance compared to the surrounding energies.

\subsection{Monolyer Graphene Diffraction}

Figure~\ref{E0=40} shows the time evolution of an electron wave packet with $E_0 = 40$ eV. The wave packet, initially propagating in the $z$ direction, is split into four components by the monolayer graphene. First-order diffraction by graphene occurs under the condition $E_0 \geq E_{\vb*{g}_1}$. Six first-order diffracted waves corresponding to the reciprocal lattice vectors
$\vb*{g}_1 = \frac{4\pi}{\sqrt{3}a}(\pm 1,0,0)$ or $\frac{4\pi}{\sqrt{3}a}(\pm 1/2,\pm \sqrt{3}/2,0)$ are generated.
An interference pattern emerges due to the interaction between these diffracted waves, resulting from the periodic boundary conditions in the $x$ and $y$ directions. For the transmitted wave packet, the zero-order diffracted wave continues propagating in the $z$ direction, while the first-order diffracted waves propagate at a diffraction angle $\theta$. 
The interference wave propagates more slowly in the $z$ direction compared to the zero-order diffracted wave.
This reduction in the $z$ component of velocity is attributed to the first-order diffracted waves possessing a velocity component $\vb*{v}_{\vb*{g}_1}$ in the $x$-$y$ plane.

Figure~\ref{E0=33} presents the EEDs of the zero-order and first-order diffracted waves after transmission through monolayer graphene with $E_0 = 33$ eV and $\alpha_0 = 40$ \AA. The EED of the zero-order diffracted wave remains unchanged from its initial state before transmission. In contrast, the EED of the first-order diffracted wave exhibits a distinct energy threshold behavior; specifically, the first-order diffracted wave intensity is zero for energies below $33.14$ eV. This occurs because first-order diffraction is only possible for electrons with energies $E > 33.14$ eV. This energy threshold leads to a narrower energy distribution for the first-order diffracted wave. The FWHM of the EED for the first-order diffracted wave, measured at $0.26$ eV,
is significantly smaller than that of the zero-order diffracted wave, which is $1.4$ eV. These simulation results suggest that highly monochromatic electrons can be obtained from monolayer graphene by utilizing the first-order diffracted wave at $E_0=33$ eV. However, the first-order diffracted wave propagates almost perpendicularly to the $z$ axis, traveling along the graphene surface. This directional constraint makes it challenging to utilize the emitted electrons effectively.  

\begin{figure*}[tb]
  \vspace{0pt}
  \centering
  \includegraphics[width=10cm]{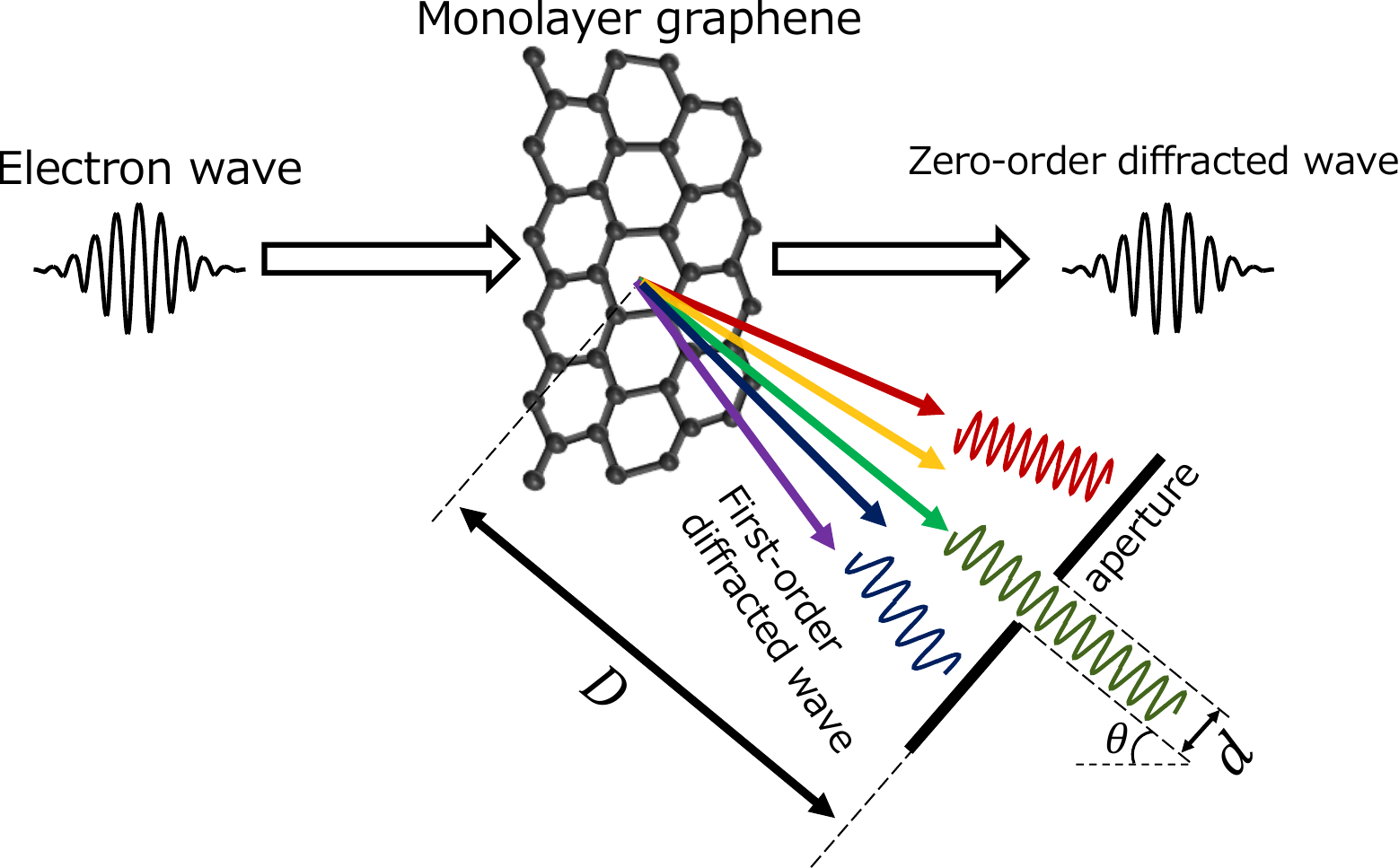}
  \caption{Schematic diagram of the method to realize highly monochromatic electron emission using a small aperture. $D$ is distance between the monolayer graphene and the aperture, $d$ is diameter of the aperture and $\theta$ is the diffraction angle.}
  \vspace{0pt}
  \label{deffraction_image}
\end{figure*}

\begin{figure}[tb]
  \vspace{0pt}
  \centering
  \includegraphics[width=7cm]{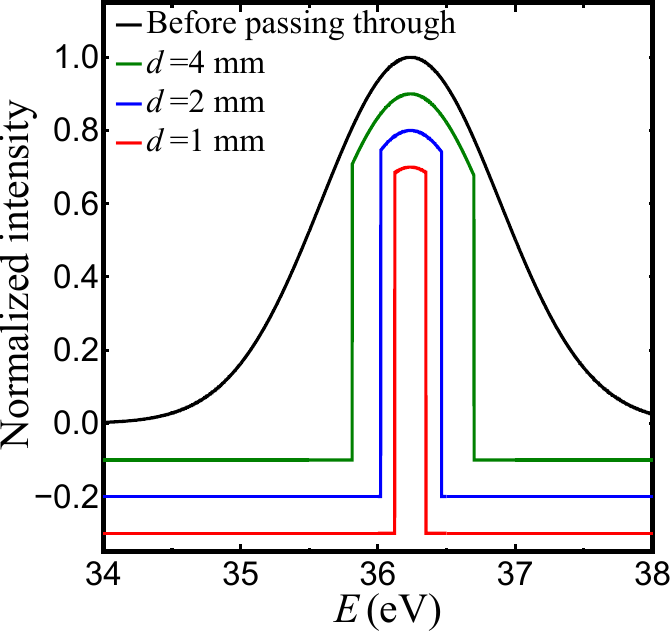}
  \caption{EEDs of the first-order diffracted electron wave 
  passing through apertures of various diameters. 
  The normalized intensities are plotted with vertically shift for easy recognition. 
  $D$ = 100\ mm and $\theta = 73^\circ$.
  The FWHM of the EED before passing through is set to 1.53\ eV.
  A narrow energy spread of $\Delta E = 0.22\,$eV is achieved by using the aperture with diameter $d = $1\,mm.
  }
  \vspace{0pt}
  \label{deffraction_d421}
\end{figure}

The wavenumber vector of the first-order diffracted wave, $\vb*{k}_{\vb*{g}_1}$ can be expressed as 
\begin{equation}
  \vb*{k}_{\vb*{g}_1} = \vb*{k}_{\vb*{g}_1\perp} + \vb*{g}_1 ,
\end{equation}
where $\vb*{k}_{\vb*{g}_1\perp}$ is the component along the $z$-direction.
The diffraction angle satisfies the following equation
\begin{equation}
    \sin \theta = \frac{\abs{\vb*{g}_1}}{\abs{\vb*{k}_{\vb*{g}_1}}} = \frac{\sqrt{E_{\vb*{g}_1}}}{\sqrt{E}}.
    \label{eq:diffraction}
\end{equation}
Note that diffraction angle $\theta$ and $E$ have a one-to-one relationship. When the electron energy $E$ is specified, the diffraction angle $\theta$ of the first-order diffracted wave can be fixed. On the contrary, the energy $E$ can be fixed from the diffraction angle $\theta$. This implies that electrons with different energies propagate in different directions. As a result, restricting the angular components of the first-order diffracted wave effectively limits the energy components of the electron beam, which forms the basis for an energy spectroscopy technique utilizing graphene. Unlike electrons propagating in the direction perpendicular to the $z$ axis, this technique is applicable not only to electrons with $E \simeq $33\,eV but also to electrons with energies greater than 33\,eV. Thus, highly monochromatic electron emission can be realized experimentally by using the first-order diffracted electron wave passing through a small aperture to limit the angular components.

Figure~\ref{deffraction_image} shows a schematic diagram of the method to achieve highly monochromatic electron emission using a small aperture. The monochromaticity of the electron beam improves with a smaller aperture diameter or an increased distance between the graphene and the aperture. Additionally, lower electron energies lead to more monochromatic electrons, as, at lower energies, corresponding to larger diffraction angles, $\abs{dE/d\theta}$ decreases, as derived from  Eq.~(\ref{eq:diffraction}). 
The viewing angle $\Delta \theta$ of an aperture with diameter $d$, placed at distance $D$, is given by 
$\Delta \theta= 2\atan (d/2D)$. Figure~\ref{deffraction_d421} shows the EEDs of the first-order diffracted electron waves passing through apertures of various diameters, with $D$ = 100\ mm, $\theta = 73^\circ$. For example, if the parameters are 
$\theta = 73^\circ$, $D$ = 100\ mm, $d$ = 1\ mm, the electron energy is restricted to the range  $36.13$--$36.35 \,$eV, which results in a highly monochromatic electron emission with a narrow energy spread of $\Delta E = 0.22\,$eV.


\subsection{Multilayer Graphene Diffraction}

\begin{figure}[tb]
  \vspace{0pt}
  \centering
  \includegraphics[width=8cm]{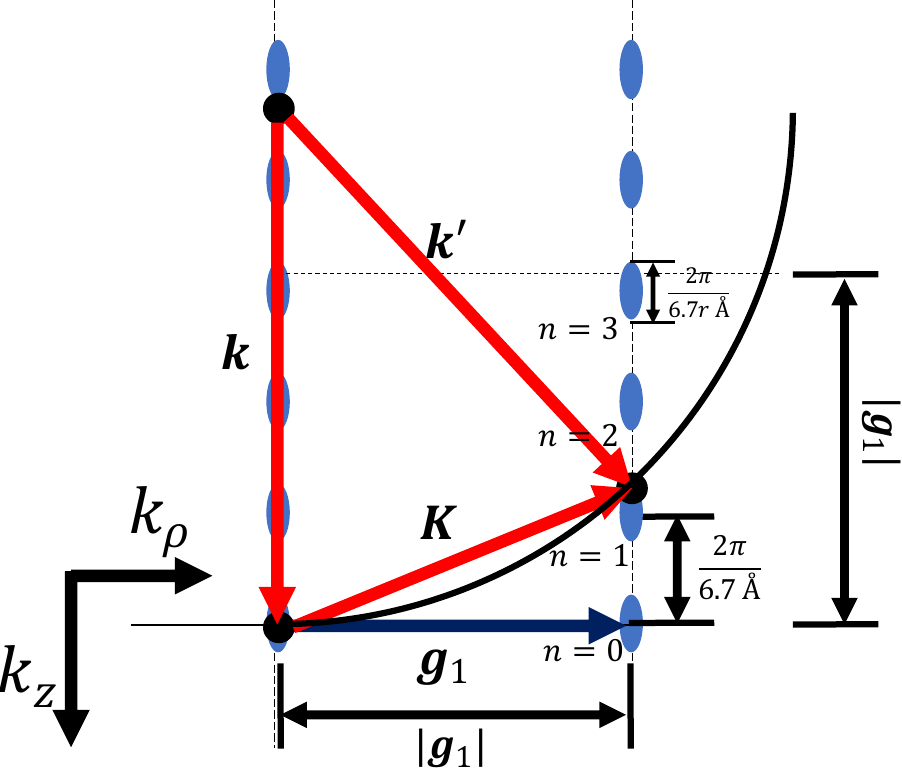}
  \caption{Reciprocal lattice points of ABA-stacked graphene.
  The spacing between the lattice points in the $k_z$ direction is $\abs{\vb*{b}_3}$.
  The spread of the points is $\abs{\vb*{b}_3}/L_3$.
  $\vb*{k_\rho} = (\vb*{k}_x,\vb*{k}_y)$.}
  \vspace{0pt}
  \label{kz_maluti}
\end{figure}

\begin{figure}[tb]
  \vspace{0pt}
  \centering
  \includegraphics[width=8cm]{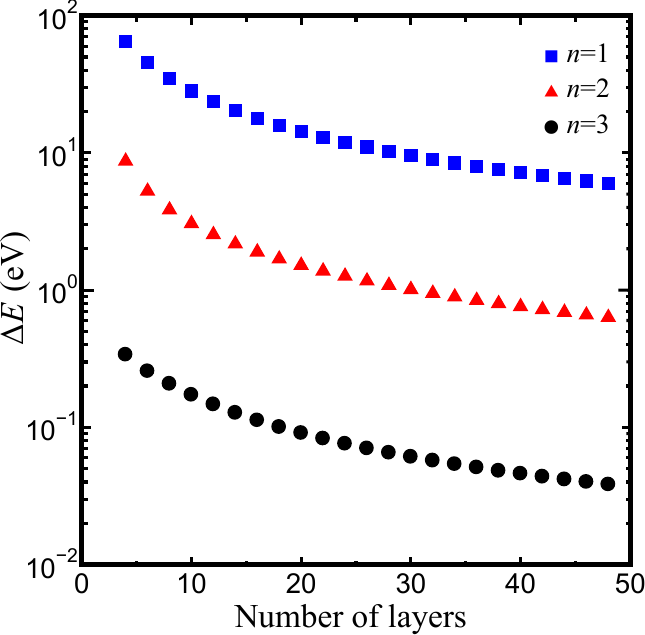}
  \caption{Energy FWHM of $\abs{f_{cr}}^2$ as a function of the number of graphene layers.
  $n$ means the position of reciprocal lattice point shown in Fig.~\ref{kz_maluti}.
  }
  \vspace{0pt}
  \label{multi}
\end{figure}

The scattered intensity is proportional to
\begin{equation}
  \abs{f_{cr}}^2 = \abs{\sum_i^N e^{-i\vb*{K}\cdot \vb*{R}_i}}^2 \abs{\sum_j e^{-i\vb*{K}\cdot \vb*{r}_j}}^2 ,
\end{equation}
where $\vb*{R}_i$ moves on all the primitive lattice points, $N=L_1 \times L_2 \times L_3$, and $L_1,L_2,L_3$ are the numbers of primitive unit cells in the $\vb*{a}_1,\vb*{a}_2,\vb*{a}_3$ directions, respectively.
$\vb*{r}_j$ represents the displacement vector in the unit cell (with $j$ labeling the atoms), and 
\begin{equation}
  \vb*{K} = \vb*{k_g} -\vb*{k}
\end{equation}
is the scattering vector\cite{Kittel, Neil}, where $\vb*{k_g}$ and $\vb*{k}$ are the wavenumbers of the diffracted and incident electron waves, respectively. Since periodic boundary conditions are assumed for the unit cell, 
\textit{i.e.} $L_1$ and $L_2$ are infinite, the reciprocal lattice points do not broaden in the $k_x$-$k_y$ plane.
The $L_3$ dimension is finite and determined by the number of graphene layers. The Bragg scattering rod is periodically segmented and converges to points as the number of graphene layers increases. The spread of the reciprocal lattice points is about $\abs{\vb*{b}_3}/L_3$. This phenomenon arises due to the crystal structure perpendicular to the graphene layer.  

Figure~\ref{kz_maluti} shows a schematic of the reciprocal lattice points of multilayer graphene. 
When discussing the transmitted electron wave, the positive integer $n$ in Fig.~\ref{kz_maluti} defined by $\vb*{K}\cdot\vb*{a}_3=2\pi n$, is restricted to $n=0,1,2,3$, because of the condition $\abs{\vb*{g}_1}>\vb*{K}\cdot \frac{\vb*{a}_3}{\abs{\vb*{a}_3}}$. Figure~\ref{multi} shows the energy FWHM of $\abs{f_{cr}}^2$ as a function of the number of graphene layers. The FWHM of $\abs{f_{cr}}^2$ decreases as the number of graphene layers increases or as $n$ increases, as shown in Fig.~\ref{multi}. A reduction in the FWHM of $\abs{f_{cr}}^2$ implies a limitation in the propagation angle of the diffracted electron waves. Thus, multilayer graphene works as a small aperture to limit angular components, as shown in Fig.~\ref{deffraction_image}.
The monochromaticity of the first-order diffracted wave improves without the need for an aperture if the appropriate incident electron energy and number of graphene layers are used.

\section{Conclusion}

We have calculated the effects of multiple reflections and diffraction of electrons by graphene. 
Interlayer interference in graphene broadens the energy spread of the electron beam for $E_0 = 13.4\,$eV.
It is shown that highly monochromatic electron emission 
can be achieved by limiting the propagation angle of the diffracted electron waves using a small aperture.
When the diffraction angle $\theta$ and the viewing angle of a small aperture $\Delta \theta$ are 
assumed to be $73^\circ$ and $\sim 0.6^\circ$, respectively, 
the energy spread $\Delta E$ is $\sim 0.22$ eV.  
The Bragg scattering rod is segmented and converges to points as the number of graphene layers increases. 
This suggests that the monochromaticity of the first-order diffracted wave improves as the number of graphene layers increases. Since the parameters used to estimate monochromaticity are realistic for actual experiments, 
experimental verification and the realization of highly monochromatic electron emission are expected.  

%
%

\section*{Acknowledgments}
{The authors would like to thank Mr. D. Terakado for the fruitful discussion. 
This work was partially supported by JSPS KAKENHI Grant Number 22H01498, 24K00954.
}

%
%
%


\nocite{*}
\bibliographystyle{unsrt}

\footnotesize	

\begin{thebibliography}{10}

\bibitem{MIM1}
Toshiaki Kusunoki, Mutsumi Suzuki, Masakazu Sagawa, Yoshiro Mikami, Etsuko
  Nishimura, Mitsuharu Ikeda, Tatsumi Hirano, Kazutaka Tsuji, 
``Highly efficient and long life metal--insulator--metal cathodes'', 
J. Vac. Sci. Technol. B \textbf{30}, 041202, 2012.

\bibitem{MIM2}
Mutsumi Suzuki, Masakazu Sagawa, Toshiaki Kusunoki, Etsuko Nishimura, Mitsuharu
  Ikeda, Kazutaka Tsuji, 
``Enhancing electron-emission efficiency of mim tunneling cathodes by
  reducing insulator trap density'', 
IEEE Trans. Electron Devices \textbf{59}, 2256--2262, 2012.

\bibitem{shimawaki}
Hidetaka Shimawaki, Yochiro Neo, Hidenori Mimura, Katsuhisa Murakami, Fujio
  Wakaya, Mikio Takai, 
``Improvement of emission efficiency of nanocrystalline silicon planar
  cathodes'', 
J. Vac. Sci. Technol. B \textbf{26}, 864--867, 2008.

\bibitem{MIM0}
CA~Mead, 
``Operation of tunnel-emission devices'', 
J. Appl. Phys. \textbf{32}, 646--652, 1961.

\bibitem{yokoo1993}
Kuniyoshi Yokoo, Hiroshi Tanaka, Shinji Sato, Junichi Murota, Shoichi Ono, 
``Emission characteristics of metal-oxide-semiconductor electron
  tunneling cathode'', 
J. Vac. Sci. Technol., B \textbf{11}, 429--432, 1993.

\bibitem{yokoo1994}
Kuniyoshi Yokoo, Shinji Sato, Gen Koshita, Isato Amano, Junichi Murota,
  Shoich Ono, 
``Energy distribution of tunneling emission from si-gate
  metal--oxide--semiconductor cathode'', 
J. Vac. Sci. Technol. B \textbf{12}, 801--805, 1994.

\bibitem{yokoo1996}
Kuniyoshi Yokoo, Gen Koshita, Satoru Hanzawa, Yoshiaki Abe, Yoichiro Neo, 
``Experiments of highly emissive metal--oxide--semiconductor electron
  tunneling cathode'', 
J. Vac. Sci. Technol. B \textbf{14}, 2096--2099, 1996.

\bibitem{yokoo2006}
Hidenori Mimura, Y~Neo, H~Shimawaki, Y~Abe, K~Tahara, K~Yokoo, 
``Improvement of the emission current from a cesiated
  metal-oxide-semiconductor cathode'', 
Appl. Phys. Lett. \textbf{88}, 123514, 2006.

\bibitem{fowler}
Ralph~Howard Fowler, Lothar Nordheim, 
``Electron emission in intense electric fields'', 
Proc. R. Soc. London, Ser. A \textbf{119}, 173--181, 1928.

\bibitem{nordhiem}
LW~Nordhiem, 
``The effect of the image force on the emission and reflexion of
  electrons by metals'', 
Proc. R. Soc. London, Ser. A \textbf{121}, 626--639, 1928.

\bibitem{100}
Katsuhisa Murakami, Shunsuke Tanaka, Akira Miyashita, Masayoshi Nagao,
  Yoshihiro Nemoto, Masaki Takeguchi, Junichi Fujita, 
``Graphene-oxide-semiconductor planar-type electron emission device'', 
Appl. Phys. Lett. \textbf{108}, 083506, 2016.

\bibitem{321}
Katsuhisa Murakami, Joji Miyaji, Ryo Furuya, Manabu Adachi, Masayoshi Nagao,
  Yoichiro Neo, Yoshinori Takao, Yoichi Yamada, Masahiro Sasaki, Hidenori
  Mimura, 
``High-performance planar-type electron source based on a
  graphene-oxide-semiconductor structure'', 
Appl. Phys. Lett. \textbf{114}, 213501, 2019.

\bibitem{furuya}
Ryo Furuya, Yoshinori Takao, Masayoshi Nagao, Katsuhisa Murakami, 
``Low-power-consumption, high-current-density, and propellantless
  cathode using graphene-oxide-semiconductor structure array'', 
Acta Astronaut. \textbf{174}, 48--54, 2020.

\bibitem{485}
Katsuhisa Murakami, Manabu Adachi, Joji Miyaji, Ryo Furuya, Masayoshi Nagao,
  Yoichi Yamada, Yoichiro Neo, Yoshinori Takao, Masahiro Sasaki, Hidenori
  Mimura, 
``Mechanism of highly efficient electron emission from a
  graphene/oxide/semiconductor structure'', 
ACS Appl. Electron. Mater. \textbf{2}, 2265--2273, 2020.

\bibitem{igari}
Tomoya Igari, Masayoshi Nagao, Kazutaka Mitsuishi, Masahiro Sasaki, Yoichi
  Yamada, Katsuhisa Murakami, 
``Origin of monochromatic electron emission from planar-type
  graphene/h-bn/n-si devices'', 
Phys. Rev. Appl. \textbf{15}, 014044, 2021.

\bibitem{Ta03}
Agnes Bogner, P-H Jouneau, Gilbert Thollet, D~Basset, Catherine Gauthier, 
``A history of scanning electron microscopy developments: Towards
  `wet-stem' imaging'', 
Micron \textbf{38}, 390--401, 2007.

\bibitem{castro2009electronic}
Antonio~H Castro~Neto, Francisco Guinea, Nuno~MR Peres, 
Kostya~S Novoselov,  Andre~K Geim, 
``The electronic properties of graphene'', 
Rev. Mod. Phys. \textbf{81}, 109--162, 2009.

\bibitem{PENG1999625}
L.-M. Peng, 
``Electron atomic scattering factors and scattering potentials of
  crystals'', 
Micron \textbf{30}, 625--648, 1999.

\bibitem{Wicki}
Flavio Wicki, Jean-Nicolas Longchamp, Tatiana Latychevskaia, Conrad Escher, Hans-Werner Fink, 
``Mapping unoccupied electronic states of freestanding graphene by
  angle-resolved low-energy electron transmission'', 
Phys. Rev. B \textbf{94}, 075424, 2016.

\bibitem{LATYCHEVSKAIA201946}
Tatiana Latychevskaia, Conrad Escher, Hans-Werner Fink.
``Moiré structures in twisted bilayer graphene studied by transmission
  electron microscopy'', 
Ultramicroscopy \textbf{197}, 46--52, 2019.

\bibitem{kavitha2016graphene}
MK~Kavitha, Manu Jaiswal, 
``Graphene: A review of optical properties and photonic applications'', 
Asian J. Phys. \textbf{25}, 809--831, 2016.

\bibitem{yang2018structure}
Gao Yang, Lihua Li, Wing~Bun Lee, Man~Cheung Ng, 
``Structure of graphene and its disorders: a review'', 
Sci. Technol. Adv. Mater. \textbf{19}, 613--648, 2018.

\bibitem{mccann2012electronic}
Edward McCann, 
\textit{Electronic Properties of Monolayer and Bilayer Graphene}, pages
  237--275, 
Springer, Berlin, Heidelberg, 2012.

\bibitem{McCann_2013}
Edward McCann, Mikito Koshino.
``The electronic properties of bilayer graphene'', 
Rep. Prog. Phys. \textbf{76}, 056503, 2013.

\bibitem{meyer2007roughness}
Jannik~C Meyer, AK~Geim, MI~Katsnelson, KS~Novoselov, D~Obergfell, S~Roth,
  C~Girit, A~Zettl, 
``On the roughness of single-and bi-layer graphene membranes'', 
Solid State Commun. \textbf{143}, 101--109, 2007.

\bibitem{meyer2007structure}
Jannik~C Meyer, Andre~K Geim, Mikhail~I Katsnelson, Konstantin~S Novoselov,
  Tim~J Booth, Siegmar Roth, 
``The structure of suspended graphene sheets'', 
Nature \textbf{446}, 60--63, 2007.

\bibitem{sung2019stacking}
Suk~Hyun Sung, Noah Schnitzer, Lola Brown, Jiwoong Park, Robert Hovden, 
``Stacking, strain, and twist in 2d materials quantified by 3d electron
  diffraction'', 
Phys. Rev. Mater. \textbf{3},  064003, 2019.

\bibitem{ABA1}
Henry~Solomon Lipson, AR~Stokes, 
``The structure of graphite'', 
Proc. R. Soc. London, Ser. A, Math. Phys. Sci. \textbf{181}, 101--105, 1942.

\bibitem{ABA2}
Masato Aoki , Hiroshi Amawashi, 
``Dependence of band structures on stacking and field in layered
  graphene'', 
Solid State Commun. \textbf{142}, 123--127, 2007.

\bibitem{Gray2009CrystalSO}
D.~Gray, Adam~Nykoruk McCaughan, Bhaskar Mookerji, 
``Crystal structure of graphite, graphene and silicon'', 
Physics for solid state applications, 2009, \\
\newblock https://api.semanticscholar.org/CorpusID:14384870.

\bibitem{CAP1}
David~E Manolopoulos, 
``Derivation and reflection properties of a transmission-free absorbing
  potential'', 
J. Chem. Phys., \textbf{117}, 9552--9559, 2002.

\bibitem{CAP2}
JG~Muga, JP~Palao, B~Navarro, IL~Egusquiza, 
``Complex absorbing potentials'', 
Phys. Rep. \textbf{395}, 357--426, 2004.

\bibitem{CAP3}
Tomas Gonzalez-Lezana, Edward~J Rackham, David~E Manolopoulos, 
``Quantum reactive scattering with a transmission-free absorbing
  potential'', 
J. Chem. Phys. \textbf{120}, 2247--2254, 2004.

\bibitem{QE}
Paolo Giannozzi, Stefano Baroni, Nicola Bonini, Matteo Calandra, Roberto Car,
  Carlo Cavazzoni, Davide Ceresoli, Guido~L Chiarotti, Matteo Cococcioni,
  Ismaila Dabo, et~al, 
``Quantum espresso: a modular and open-source software project for
  quantum simulations of materials'', 
J. Phys.: Condens. Matter. \textbf{21}, 395502, 2009.

\bibitem{simulation_yan}
Jia-An Yan, JA~Driscoll, BK~Wyatt, K~Varga, ST~Pantelides, 
``Time-domain simulation of electron diffraction in crystals'', 
Phys. Rev. B \textbf{84}, 224117, 2011.

\bibitem{Takao1}
Takao Koichi, Shogo Kawashima, Satoshi Abo, Fujio Wakaya, Masayoshi Nagao, Katsuhisa Murakami, 
``Simulation of electron transmission through graphene with inelastic
  scattering'', 
e-J. Surf. Sci. Nanotechnol. \textbf{22}, 157--161, 2024.

\bibitem{Miyauchi}
Hironari Miyauchi, Yoshihiro Ueda, Yasumitsu Suzuki, Kazuyuki Watanabe, 
``Electron transmission through bilayer graphene: A time-dependent
  first-principles study'', 
Phys. Rev. B \textbf{95}, 125425, 2017.

\bibitem{LATYCHEVSKAIA2023113807}
Tatiana Latychevskaia, 
``Coherent imaging with low-energy electrons, quantitative analysis'', 
Ultramicroscopy \textbf{253}, 113807, 2023.

\bibitem{Longchamp}
Jean-Nicolas Longchamp, Tatiana Latychevskaia, Conrad Escher, Hans-Werner
  Fink, 
``Low-energy electron transmission imaging of clusters on free-standing
  graphene'', 
Appl. Phys. Lett. \textbf{101}, 113117, 2012.

\bibitem{Kittel}
Charles Kittel.
\textit{Introduction to Solid State Physics}, 
John Wiley \& Sons, 8 edition, 2004.

\bibitem{Neil}
N.W. Ashcroft, N.D. Mermin, D.~Wei, 
\textit{Solid State Physics}, 
Cengage Learning, Singapore, 2016.

\end{thebibliography}


\end{document}